\documentclass[a4paper,fleqn,usenatbib]{mnras}

\usepackage{graphicx}
\usepackage{url}
\usepackage{enumitem}
\usepackage{amssymb}
\usepackage{amsmath}
\usepackage{rotating}
\usepackage{pdflscape}

\title[Metallicity of M~2]{The iron dispersion of the globular cluster M~2, revised}
\author[C.~Lardo, A.~Mucciarelli \& N.~Bastian]{C. Lardo$^{1}$\thanks{E-mail:
C.Lardo@ljmu.ac.uk}, A. Mucciarelli $^{2}$, and N. Bastian$^{1}$\\
$^{1}$ Astrophysics Research Institute, Liverpool John Moores University, 146 Brownlow Hill, Liverpool, L3 5RF, UK\\
$^{2}$ Department of Physics and Astronomy, University of Bologna, Viale Berti Pichat 6/2, I-40127 Bologna, Italy}

\date{Accepted XX . Received XX; in original form XX}
\pubyear{2015}

\begin{document}
\label{firstpage}
\pagerange{\pageref{firstpage}--\pageref{lastpage}}
\maketitle

\begin{abstract}
M2 has been claimed to posses three distinct stellar components that are enhanced in iron relative to each other.
We use equivalent width measurements from 14 red giant branch stars from which Yong et al. detect a 
$\sim$0.8 dex wide, trimodal iron distribution to redetermine the metallicity of the cluster. 
In contrast to Yong et al., which derive atmospheric parameters following only the classical spectroscopic approach, 
we perform the chemical analysis using three different methods to constrain effective temperatures and surface gravities.
When atmospheric parameters are derived spectroscopically, we measure
 a trimodal metallicity distribution, that well resembles that by Yong et al.
We find that the metallicity distribution from  Fe\,{\sc ii} lines strongly differs from the distribution obtained from Fe\,{\sc i} features when photometric gravities are adopted. The Fe\,{\sc i} distribution  mimics the metallicity distribution obtained 
using spectroscopic parameters, while  the Fe\,{\sc ii} shows the presence of only two stellar groups with metallicity [Fe/H]$\simeq$--1.5 and --1.1 dex, which are internally homogeneous in iron. 
This finding, when coupled with the high-resolution photometric evidence, demonstrates that M~2 is composed by a dominant population
($\sim$99\%) homogeneous in iron and a minority component ($\sim$1\%) enriched in iron with respect to the main cluster population.

\end{abstract}

\begin{keywords}
stars: abundances -- stars: atmospheres -- stars: evolution -- stars: Population II --globular clusters: individual: NGC 7089
\end{keywords}

\section{Introduction}\label{introduzione}
Most of the globular clusters (GCs) surveyed so far
show large internal variations in 
the abundances of light elements (C, N, O, Na, and Al; i.e.,  \citealp{kraft94}, 
\citealp{gratton04}, \citealp{carretta2009}, \citealp{car09b}).
In contrast, only a small subset of the cluster population is characterised by star-to-star 
variations in their heavy-element content \citep[e.g.][]{grattonREV}.

Among clusters with intrinsic spreads in heavy elements, of great interest are 
those characterised by a dispersion in their slow ($s$)-capture element abundance content
(namely; NGC~1851, M~22, NGC~362, M~2, NGC~5286, and M~19).
In these GCs, stars are clustered in two groups with different $s$-process element
content. Stars with higher $s$-process element content in M~22 \citep{marino09,marino11}, M~2 \citep{yong14}, 
NGC~5286 \citep{marino15}, NGC~1851 (\citealp{yong08,carretta1851}, but see \citealp{villanova10}), and 
M~19 \citep{johnson15} are also believed to be more metal-rich than stars with no $s$-process element overabundance, 
the characteristic range in iron being $\simeq$ 0.2--0.4 dex.
Also, the $s$-process bimodality is associated to the photometric 
split in the sub-giant branch (SGB) in $V, V-I$ colour-magnitude diagrams (CMDs; \citealp{milone08,piotto09,piotto12}) and to the multimodal
red giant branch (RGB) when a combination of colours, including the $U$ filter is used to construct the CMD (\citealp{marino12,carretta1851,lardoM2,lardo13, yong14,carretta362, marino15}).

The presence of internal variations in iron would suggest that clusters with Fe spreads 
have been able to keep high-energy supernova (SN) ejecta.
This in turn indicates that M~22, M~2, NGC~5286, NGC~1851, and {M~19} 
 were much more massive at their birth, possibly being nuclei 
of dwarfs disrupted by Milky Way tidal fields (e.g. \citealp{marino15}, but see also \citealp{pfeffer14}), as SN ejecta 
are too energetic to be retained by less massive systems like 
Galactic GCs,  which have typical masses less than or equal to a few times 10$^{5}$ M$_{\odot}$ \citep{baumgardt08}.
In this scenario GCs showing intrinsic metallicity variations may be considered as former Milky Way satellites, 
and this assumption greatly impacts on our understanding of how the Galaxy formed and evolved.
Therefore, it is crucial to assess whether the observed star-to-star variations are {\em genuine} metallicity dispersions.

Iron abundances critically depend on the choice of atmospheric parameters
adopted in the spectroscopic analysis.
When measuring abundances, two main approaches are routinely used to constrain 
surface gravities from stellar spectra. 
When spectra have very good quality; i.e. high signal-to-noise ratio (SNR), wide spectral coverage, 
relatively large ($\sim$10-20) number of  observed Fe\,{\sc ii} lines in the covered spectral range, one commonly sets 
the surface gravity from the ionisation equilibrium of Fe; i.e. in a fashion that the same iron abundance is provided by 
both Fe\,{\sc i} and Fe\,{\sc ii} lines. When this approach is not  feasible, i.e. in case of spectra with 
relatively low  SNR and/or resolution, very low-metallicity, and/or 
small spectral coverage; one has to rely 
on gravities estimated from photometric data, such as those derived
using colour-temperature and colour-bolometric correction calibrations or 
isochrone fitting \citep[e.g.][]{carretta2009,car09b}.

Very recently, \citealp{mucciarelliM22} (Mu15) found that the  [Fe\,{\sc ii}/H] distribution 
of M~22 stars is very narrow when the surface gravities are estimated from photometry.
Conversely, when gravities are derived from the spectra by imposing the ionisation equilibrium
between [Fe\,{\sc i}/H]  and [Fe\,{\sc ii}/H] lines, the iron distribution is $\simeq$ 0.5 dex wide,
in agreement with previous results \citep{marino09,marino11}. 
Naively, spectroscopic surface gravities appear more reliable than photometric ones, because they are
quantitatively extracted from spectra and they are not estimated form an empirical calibration or stellar evolution, 
as in the case of photometric gravities. However, Mu15 find that 
spectroscopic gravities required to match [Fe\,{\sc i}/H]  and [Fe\,{\sc ii}/H] abundances lead to
{\em unphysical} stellar masses for GC giant stars, with values $\leq$ 0.5 M$_{\odot}$.
The Mu15 sample is composed by stars in the same evolutionary stage, located at the same distance and with virtually the same age;
and their spectroscopic masses are spread over a wide range ($\simeq$0.8 M$_{\sun}$).
Since it is hard to think to a physical mechanism which {\em randomly} scatter stellar masses,
Mu15 conclude that the Fe spread observed among M~22 stars is artificially produced by the method
used to constrain surface gravities.

M~2 is another cluster that has been claimed to posses an intrinsic iron spread, with possibly 
three populations at [Fe/H]=--1.7,--1.5, and --1.0 dex (\citealp{yong14}; hereafter Y14).
The metal-poor and metal-intermediate  components include stars with different $s$-process abundances
with stars at [Fe/H]$\simeq$--1.5 dex being $s$-rich (\citealp{lardo13}, Y14) with respect to metal-poor stars at [Fe/H]$\simeq$--1.7 dex. 
In contrast with other clusters showing $s$-process element bimodality, M~2 is characterised by a third, poorly populated stellar group,
which does not show any GC anti-correlations or $s$-process enrichment (Y14).

In their analysis, Y14 take advantage from the wide wavelength coverage and high SNR of their spectra
to constrain surface gravities from the ionisation balance. 
In the light of Mu15 results, we wonder whether the spread observed by Y14 is a genuine iron dispersion and
we apply the same analysis as in Mu15 to the spectroscopic sample presented by \citet{yong14} 
to re-derive iron abundances.

This article is structured as follows:
we describe the observational material and data in Section~\ref{OSSERVAZIONI}. We measure 
iron abundances in Section~\ref{gspectro}. We discuss our results in Section~\ref{DISCUSSION} and draw our conclusions in Section~\ref{CONCLUSIONI}.

\begin{table}
\setlength{\tabcolsep}{0.28cm}
\caption{M~2 stars analysed in this paper, together with the photometric information and the [Fe/H] estimates by  \citet{yong14}. 
The stars are grouped according to the classification proposed by \citet{yong14}.}
\label{PHOT_PAR}
\begin{tabular}{@{}lccccc}
\hline
ID      &        $U$     &  $B$     &    $V$    &     $I $   &  [Fe/H]  \\
	 & (mag)  & (mag)  &   (mag)    &  (mag)    &   (dex)      \\
          \hline
            \multicolumn{6}{c}{ Metal-poor stars with [Fe/H]$\simeq$ --1.7 dex}\\
            \hline
   NR~37    & 15.428   &  14.717   &  13.556  &  12.267   & --1.66    \\
   NR~58   & 15.838   &  14.832   &  13.596  &  12.243   & --1.64   \\
   NR~60  & 15.747   &  14.828   &  13.620  &  12.309   & --1.75     \\
   NR~76   & 15.810   &  15.030   &  13.906  &  12.647   & --1.69     \\
   NR~99   & 15.816   &  14.950   &  13.746  &  12.433   & --1.66    \\
  NR~124    & 15.886   &  15.217   &  14.148  &  12.944   & --1.64     \\
          \hline
            \multicolumn{6}{c}{ Intermediate-metallicity stars with [Fe/H]$\simeq$ --1.5 dex} \\
            \hline
   NR~38   & 16.457   &  15.056   &  13.687  &  12.339   & --1.61  \\
   NR~47   &  --   &  14.837   &  13.534  &  12.116   & --1.42   \\
   NR~77   &  --   &  15.207   &  13.937  &  12.704   & --1.46      \\
   NR~81  & 16.318   &  15.101   &  13.821  &  12.523   & --1.55    \\
        \hline
            \multicolumn{6}{c}{ Metal-rich stars with [Fe/H]$\simeq$ --1.0 dex} \\
            \hline
      NR~132   & 16.837   &  15.534   &  14.249  &  12.880   & --0.97   \\
  NR~207   &  --   &  16.055   &  14.937  &  13.726   & --1.08 \\
  NR~254   & 16.926   &  16.151   &  15.073  &  13.878   & --0.97 \\
  NR~378   & 16.621   &  16.170   &  15.254  &  14.207   & --1.08   \\
\hline
\end{tabular}
\end{table}
\section{Observational material and EW data}\label{OSSERVAZIONI}

The original high-resolution sample analysed in Y14 is composed by 14 RGB stars selected from 
$uvby$ Str\"{o}mgren photometry by \citet{grundahl99}. Five stars (NR 37, NR 38, NR 58, NR 60 and NR 77) were observed using the Magellan Inamori Kyocera Echelle (MIKE) spectrograph \citep{bernstein03} at the Magellan Telescope on 2012 August 26. 
 MIKE spectra cover a wavelength range between $\simeq$3400  to 9000\AA, and
have a spectral resolution of R = 40 000 in the blue arm and R = 35 000 in the red arm. 

The remaining stars were observed 
with the High Dispersion Spectrograph (HDS; \citealp{noguchi02})
at the Subaru telescope. Echelle spectroscopy was obtained on 2011 August 3 for three stars (NR~76, NR~81, and NR~132).
Six additional stars (NR 47, NR 99, NR 124, NR 207, NR 254 and NR 378) were observed using HDS in classical mode on 2013 July 17. 
The HDS spectrograph was used in the StdYb setting and the 0.8\arcsec slit, 
providing spectra with typical resolution of R $\simeq$ 45 000 and a spectral coverage from $\simeq$4100 to  $\simeq$6800~\AA.
We refer to Y14 for further details on observations and data reduction.

\begin{figure}
  \centering
\includegraphics[width=0.9\columnwidth]{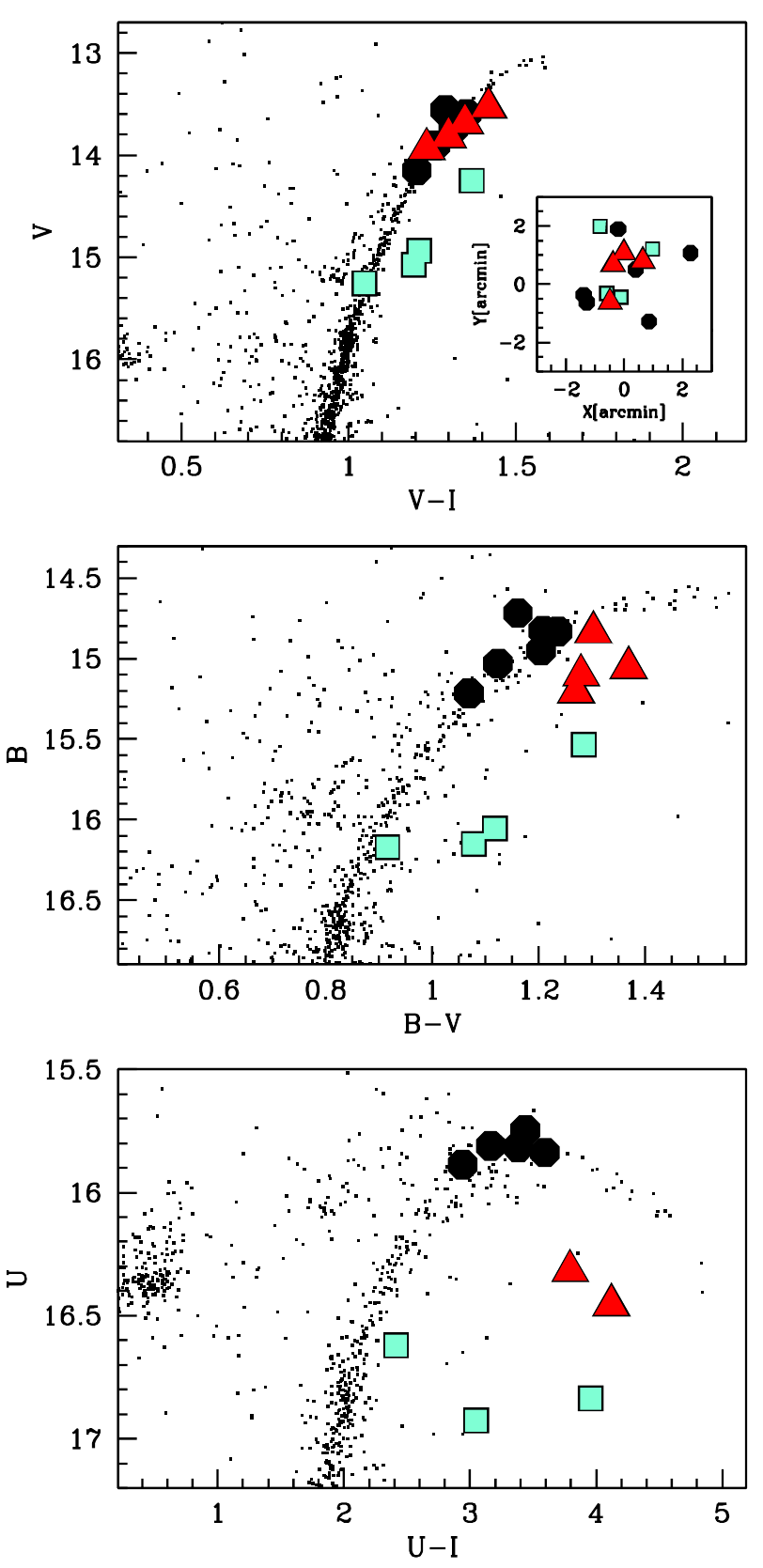}
\caption{CMDs  for $V$ versus $V-I$ (upper), $B$ versus $B-V$ (middle) and $U$ versus $U-I$ (lower) for the spectroscopic targets analysed by Y14 at high spectral resolution. Note that three stars do not have $U$ magnitude in our photometric catalogue. Therefore they are missing in the 
$U$ versus $U-I$  CMD.
The inset in the top panel shows the location of targets across the cluster.
The black symbols show metal-poor ([Fe/H]$\simeq$--1.7 dex) stars according to Y14 analysis.
The red and aqua symbols denote the metal-intermediate ([Fe/H]$\simeq$--1.5 dex) and metal-rich ([Fe/H]$\simeq$--1.0 dex) components identified by Y14, respectively. 
}
        \label{CMD}
   \end{figure}

Unpublished Wide-Field Imager (WFI) photometry from one of us was used to identify the targets.
Figure~\ref{CMD} shows the position of the spectroscopic targets in the $V$ versus $V-I$, $B$ versus $B-V$ and $U$ versus $U-I$
colour-magnitude diagrams (CMDs). Their magnitudes are listed in Table~\ref{PHOT_PAR}.
The photometric catalogue is based on archival $UBVI$ images collected at the WFI at the 2.2 m ESO-MPI telescope and used to select targets for the Gaia ESO survey calibration \citep{pancino12}. The WFI covers a total field of view of 34$\arcmin$ $\times$ 33$\arcmin$, consisting of 8, 2048 $\times$ 4096 EEV-CCDs with a pixel size of 0.238$\arcsec$. These images were pre-reduced using the {\tt IRAF} package {\tt MSCRED} \citep{valdes98}, while the stellar photometry was derived by using the {\tt DAOPHOT II} and {\tt ALLSTAR} programs \citep{stetson87,stetson92}. Details on the preproduction, calibration, and full photometric catalogues will be published elsewhere.

Figure~\ref{CMD} displays the presence of an additional RGB sequence (\citealp{lardoM2}, Y14, \citealp{milone15}).
We use the same colour code as Y14 to represent stars that have metallicity 
[Fe/H] $\simeq$ --1.7, --1.5, and --1.0 dex according to their analysis (black circles, red triangles, and aqua squares, respectively).
The metal-poor and metal-intermediate stars are 
tightly aligned along the RGB in the $V$, $V-I$  CMD of Figure~\ref{CMD}, while they are clearly separated into two sequences in the split 
RGBs observed in the $B-V$ and $U-I$ colours. 
The observed effect can be attributed to the presence of molecular CN and CH absorption in 
wavelength range covered by the $U$ and $B$ filters \citep[see also][]{milone15}.
In \citet{lardo13} we demonstrate that red RGB stars selected in the $V$, $U-V$ CMD \citep{lardoM2} are indeed richer (on average) in both 
C and N with respect to the stars located on the blue side of the RGB. 
Both the metal-poor and metal-intermediate components display the light element pattern commonly found in GCs (Y14).

Metal-rich stars (with [Fe/H] $\simeq$--1.0 dex) are located along a redder sequence which runs parallel to the RGB 
in all the CMDs of Figure~\ref{CMD} because of their higher metallicity with respect to the bulk of M~2 stars (Y14).
Metal-rich stars do not show any GC like anticorrelation or $s$-enhancement.

We refer again to Y14 for a complete discussion on cluster membership.
Briefly, all stars have colours and magnitudes consistent with being giant stars at the distance of M2 and 
seven stars have probability P=99\% to be cluster members, according the proper motion study by 
\citet{cudworth87}. However, the heliocentric radial velocity of M~2 is --5.3 $\pm$2 km s$^{-1}$ and the central velocity dispersion is 8.2 $\pm$0.6 km s$^{-1}$ \citep{harris96}. Therefore radial velocities alone cannot confirm cluster membership, 
as field stars could easily have such velocities\footnote{Nonetheless, the probability of finding a field star at [Fe/H] $\leq$ --1.5 dex
with kinematics compatible with the cluster is extremely low \citep{lardoM2}.}.

Hubble Space Telescope (HST)  photometry by \citet{milone15} shows that three metal-rich stars are located 
on a defined RGB sequence that can be followed down to the SGB and main sequence, 
supporting the case for cluster membership. 
However, from Figure~\ref{CMD} we note that star NR~378 is 
systematically bluer than the other metal-rich stars in all the CMDs of Figure~\ref{CMD}. The same is observed in Figure~1 of Y14
in $uvby$ Str\"{o}mgren colours. In Section~\ref{gspectro} we measure for this star a slightly higher metallicity than 
metal-rich stars. Its location in the CMD and its higher metallicity would indicate NR~378 as a non member, but
proper motion or parallaxes are needed to settle the case for this star.

Additionally, Figure~\ref{CMD} suggests that NR~60 could be an asymptotic giant branch (AGB) star, while NR~37 looks  slightly off 
the RGB in Figure~\ref{CMD}. To confirm this suggestion, we plotted both stars in the $V, V-I$ CMD (not shown) by \citet{sarajedini07}. We find that NR~60 is indeed an AGB star (see also Y14), while NR~37 is located on the main RGB body of the cluster in the higher precision HST photometry.

 \begin{figure*}
\includegraphics[width=1.5\columnwidth]{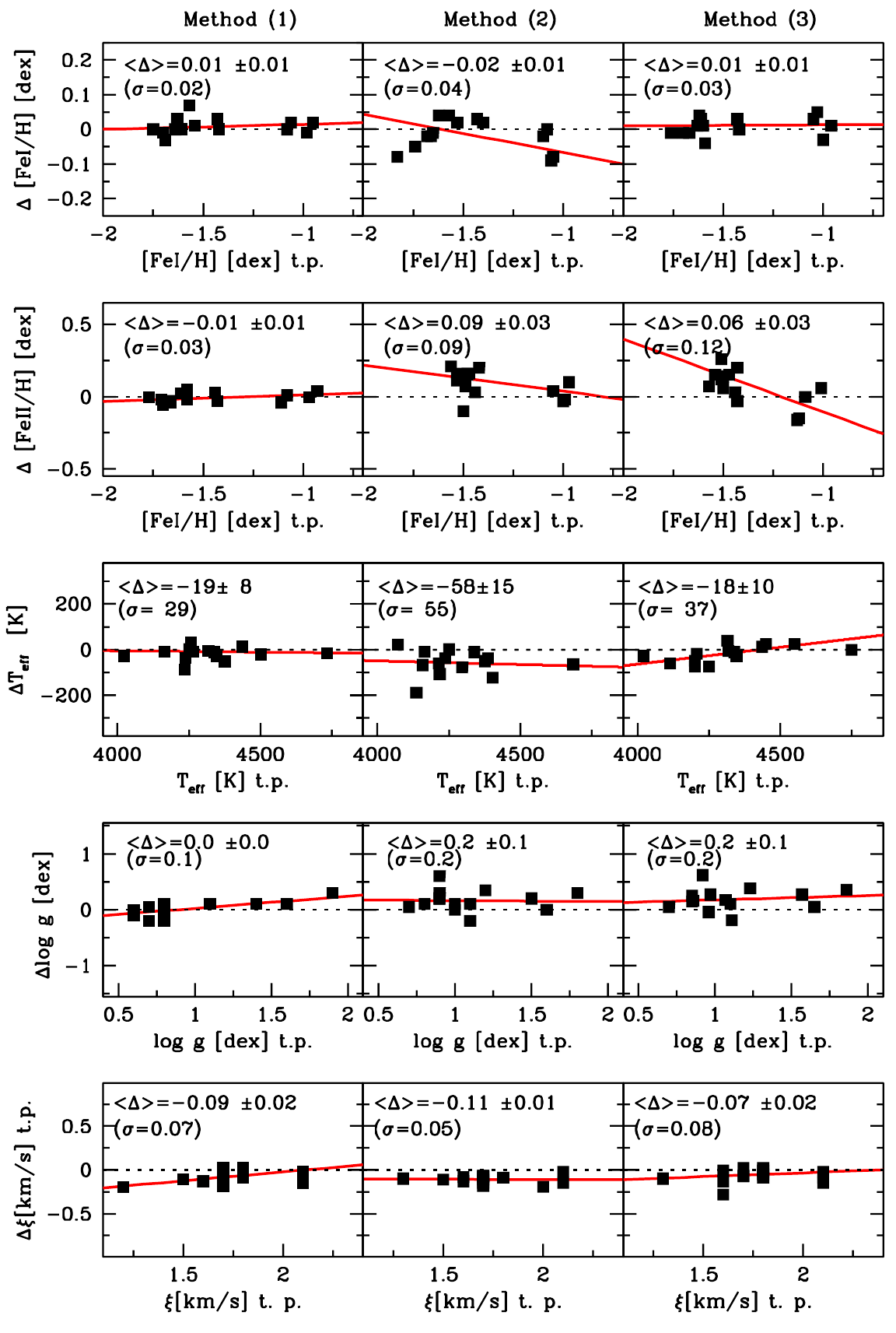}
\caption{A comparison between the atmospheric parameters adopted in our analysis and those listed in Y14 is presented. Each panel reports in the top left corner the mean difference between (from top to bottom) the [Fe\,{\sc i}/H], [Fe\,{\sc ii}/H], T$_{\rm{eff}}$, $\log$ g, and $\xi_{\rm{t}}$ measured as detailed in Section~\ref{gspectro} and those listed in Y14. Also shown the 1:1 relationships (dotted lines) and 
the weighted linear fit to the data (solid lines).  }
        \label{IRON}
   \end{figure*}

\begin{footnotesize}
\begin{table*}

\setlength{\tabcolsep}{0.10cm}

\caption{Atmospheric parameters and metallicities as derived with the methods described in Section~\ref{gspectro}.}
\label{FERRO_TAB}

\begin{tabular} {lcccccccccc}
\hline
\hline
ID     &  T$_{\rm{eff}}$  &  $\log$ g      & $\xi$          & [M/H]    & [Fe\,{\sc i}/H]  & $\delta_{\rm{int}}$ & $\delta_{\rm{par}}$  & [Fe\,{\sc ii}/H] & $\delta_{\rm{int}}$ & $\delta_{\rm{par}}$  \\
              &  (K)        &  (dex)      & (kms$^{-1}$)          &  (dex)           & (dex)  & (dex) & (dex) & (dex) & (dex)& (dex)  \\
\hline

&\multicolumn{10}{c}{ {\em (1)} Spectroscopic analysis} \\
\cline {2-11}
                   &           &          &          &            &          &                       &            &             &          &         \\
 NR~37	 &  4251$\pm$28  &  0.60$\pm$0.06 &  1.70$\pm$0.04 & --1.50  &  --1.63  &  0.01  &   0.03  &  --1.67 &  0.03 & 0.05 \\
NR~58	 &  4258$\pm$28  &  0.80$\pm$0.06 &  1.80$\pm$0.06 & --1.50  &  --1.57  &  0.01  &   0.03  &  --1.58 &  0.04 & 0.06 \\
NR~60	 &  4318$\pm$29  &  0.30$\pm$0.06 &  2.10$\pm$0.07 & --1.50  &  --1.75  &  0.01  &   0.04  &  --1.77 &  0.03 & 0.05 \\
NR~76	 &  4347$\pm$33  &  0.70$\pm$0.06 &  1.60$\pm$0.04 & --1.50  &  --1.70  &  0.01  &   0.04  &  --1.71 &  0.03 & 0.05 \\
NR~99	 &  4265$\pm$28  &  0.60$\pm$0.06 &  1.80$\pm$0.04 & --1.50  &  --1.69  &  0.01  &   0.03  &  --1.70 &  0.03 & 0.05 \\
NR~124    &  4437$\pm$46  &  0.80$\pm$0.06 &  1.80$\pm$0.06 & --1.50  &  --1.64  &  0.01  &   0.06  &  --1.66 &  0.04 & 0.06 \\
NR~38	 &  4165$\pm$44  &  0.60$\pm$0.06 &  2.10$\pm$0.08 & --1.50 &  --1.61  &  0.01  &   0.03  &  --1.61 &  0.04 & 0.07 \\
NR~47	 &  4022$\pm$80  &  0.70$\pm$0.06 &  1.70$\pm$0.10 & --1.50  &  --1.42  &  0.02  &   0.05  &  --1.43 &  0.04 & 0.13 \\
NR~77	 &  4339$\pm$78  &  1.10$\pm$0.06 &  2.10$\pm$0.14 & --1.50  &  --1.43  &  0.02  &   0.08  &  --1.44 &  0.09 & 0.09 \\
NR~81	 &  4237$\pm$49  &  0.80$\pm$0.06 &  1.70$\pm$0.06 & --1.50  &  --1.54  &  0.01  &   0.05  &  --1.58 &  0.03 & 0.07 \\
NR~132    &  4236$\pm$55  &  1.40$\pm$0.06 &  1.70$\pm$0.07 & --1.00  &  --0.98  &  0.01  &   0.03  &  --0.97 &  0.04 & 0.09 \\
NR~207    &  4375$\pm$61  &  1.40$\pm$0.06 &  1.20$\pm$0.04 & --1.00  &  --1.06  &  0.01  &   0.05  &  --1.08 &  0.04 & 0.08 \\
NR~254    &  4503$\pm$61  &  1.90$\pm$0.06 &  1.50$\pm$0.05 & --1.00  &  --0.95  &  0.01  &   0.05  &  --0.93 &  0.04 & 0.08 \\
NR~378    &  4735$\pm$42  &  1.60$\pm$0.07 &  1.70$\pm$0.08 & --1.00  &  --1.08  &  0.02  &   0.05  &  --1.11 &  0.02 & 0.05 \\
              &\multicolumn{10}{c}{{\em (2)} Photometric analysis} \\
\cline {2-11}
                                     &           &          &          &                      &             &            &            &             &          &         \\
NR~37	 &  4251$\pm$ 83 &  0.90$\pm$0.10  & 1.70$\pm$0.04 & --1.50  &  --1.62  &  0.01  &   0.07  &  --1.53 &  0.03 & 0.09 \\							  
NR~58	 &  4158$\pm$ 76 &  0.80$\pm$0.10  & 1.80$\pm$0.06 & --1.50  &  --1.66  &  0.01  &   0.06  &  --1.48 &  0.04 & 0.11 \\							  
NR~60	 &  4218$\pm$ 80 &  0.90$\pm$0.10  & 2.00$\pm$0.07 & --1.50  &  --1.83  &  0.01  &   0.09  &  --1.56 &  0.03 & 0.08 \\							  
NR~76    &  4297$\pm$ 86 &  1.00$\pm$0.10  & 1.60$\pm$0.06 & --1.50  &  --1.74  &  0.01  &   0.08  &  --1.53 &  0.03 & 0.10 \\							  
NR~99	 &  4215$\pm$ 80 &  0.90$\pm$0.10  & 1.70$\pm$0.04 & --1.50  &  --1.68  &  0.01  &   0.07  &  --1.48 &  0.04 & 0.10 \\							  
NR~124	 &  4387$\pm$ 93 &  1.20$\pm$0.10  & 1.70$\pm$0.07 & --1.50  &  --1.65  &  0.01  &   0.10  &  --1.42 &  0.04 & 0.09 \\							  
NR~38	 &  4165$\pm$ 76 &  0.90$\pm$0.10  & 2.10$\pm$0.09 & --1.50  &  --1.57  &  0.01  &   0.05  &  --1.47 &  0.04 & 0.11 \\							  
NR~47	 &  4072$\pm$ 69 &  0.70$\pm$0.10  & 1.70$\pm$0.10 & --1.50  &  --1.40  &  0.02  &   0.05  &  --1.50 &  0.04 & 0.11 \\							  
NR~77	 &  4339$\pm$ 89 &  1.10$\pm$0.10  & 2.10$\pm$0.14 & --1.50  &  --1.43  &  0.02  &   0.09  &  --1.44 &  0.09 & 0.10 \\							  
NR~81	 &  4237$\pm$ 82 &  1.00$\pm$0.10  & 1.70$\pm$0.07 & --1.50  &  --1.53  &  0.01  &   0.06  &  --1.49 &  0.03 & 0.10 \\							  
NR~132   &  4136$\pm$ 74 &  1.10$\pm$0.10  & 1.70$\pm$0.06 & --1.00  &  --1.06  &  0.01  &   0.03  &  --1.00 &  0.04 & 0.11 \\							  
NR~207   &  4375$\pm$ 92 &  1.50$\pm$0.10  & 1.30$\pm$0.06 & --1.00  &  --1.08  &  0.01  &   0.06  &  --1.05 &  0.04 & 0.11 \\							  
NR~254   &  4403$\pm$ 94 &  1.60$\pm$0.10  & 1.50$\pm$0.05 & --1.00  &  --1.05  &  0.01  &   0.07  &  --0.99 &  0.04 & 0.11 \\							  
NR~378   &  4685$\pm$115 &  1.80$\pm$0.10  & 1.60$\pm$0.08 & --1.00  &  --1.10  &  0.02  &   0.12  &  --0.97 &  0.02 & 0.08 \\							  
						  
	&\multicolumn{10}{c}{{\em (3)} Hybrid analysis} \\
\cline {2-11}
                    &           &          &          &            &          &                     &            &             &          &         \\
NR~37	 &  4201$\pm$ 30 &  0.86$\pm$0.06 &  1.70$\pm$0.04 & --1.50  &  --1.67  &  0.01  &   0.03  &  --1.50 &  0.03 & 0.06 \\						   
NR~58	 &  4208$\pm$ 30 &  0.85$\pm$0.06 &  1.80$\pm$0.06 & --1.50  &  --1.61  &  0.01  &   0.03  &  --1.51 &  0.04 & 0.06 \\						   
NR~60	 &  4318$\pm$ 32 &  0.92$\pm$0.06 &  2.10$\pm$0.07 & --1.50  &  --1.76  &  0.01  &   0.03  &  --1.51 &  0.03 & 0.05 \\						   
NR~76    &  4347$\pm$ 39 &  1.07$\pm$0.06 &  1.60$\pm$0.06 & --1.50  &  --1.70  &  0.01  &   0.04  &  --1.54 &  0.03 & 0.06 \\						   
NR~99	 &  4315$\pm$ 29 &  0.97$\pm$0.08 &  1.80$\pm$0.04 & --1.50  &  --1.62  &  0.01  &   0.03  &  --1.57 &  0.03 & 0.06 \\						   
NR~124	 &  4437$\pm$ 51 &  1.23$\pm$0.07 &  1.80$\pm$0.07 & --1.50  &  --1.63  &  0.01  &   0.06  &  --1.47 &  0.04 & 0.06 \\						   
NR~38	 &  4115$\pm$ 48 &  0.85$\pm$0.06 &  2.10$\pm$0.09 & --1.50  &  --1.60  &  0.01  &   0.03  &  --1.43 &  0.04 & 0.09 \\						   
NR~47	 &  4022$\pm$ 80 &  0.70$\pm$0.06 &  1.70$\pm$0.10 & --1.50  &  --1.42  &  0.02  &   0.05  &  --1.43 &  0.04 & 0.13 \\						   
NR~77	 &  4339$\pm$ 78 &  1.10$\pm$0.06 &  2.10$\pm$0.14 & --1.50 &  --1.43  &  0.02  &   0.08  &  --1.44 &  0.09 & 0.09 \\						   
NR~81	 &  4200$\pm$ 51 &  0.96$\pm$0.06 &  1.80$\pm$0.07 & --1.50  &  --1.59  &  0.01  &   0.04  &  --1.50 &  0.04 & 0.08 \\						   
NR~132   &  4250$\pm$ 30 &  1.11$\pm$0.06 &  1.60$\pm$0.09 & --1.00  &  --1.00  &  0.01  &   0.02  &  --1.13 &  0.03 & 0.07 \\						   
NR~207   &  4450$\pm$ 41 &  1.57$\pm$0.06 &  1.30$\pm$0.04 & --1.00  &  --1.03  &  0.01  &   0.04  &  --1.09 &  0.05 & 0.06 \\						   
NR~254   &  4550$\pm$ 51 &  1.65$\pm$0.06 &  1.60$\pm$0.05 & --1.00 &  --0.96  &  0.01  &   0.05  &  --1.12 &  0.04 & 0.06 \\						   
NR~378   &  4750$\pm$ 52 &  1.86$\pm$0.07 &  1.70$\pm$0.10 & --1.00  &  --1.05  &  0.02  &   0.06  &  --1.01 &  0.02 & 0.06 \\

\hline
\end{tabular}
\flushleft{{\bf Notes:} [M/H] refers to the metallicity used to generate the model atmosphere.}
\end{table*}
\end{footnotesize}

 \section{Iron abundances}\label{gspectro}
 In the following, to be consistent with the spectroscopic analysis presented in Y14, we adopt 
both their equivalent widths (EW) measurements and their atomic line list (see Table~3 in Y14).
The main literature sources for Fe\,{\sc i} and  Fe\,{\sc i} atomic data are from the Oxford group, including  \citet{blackwell79a,blackwell79b,blackwell80,blackwell86,blackwell95}, \citet{biemont91}, and \citet{gratton03}.
We refer to Y14 for additional  details on how the atomic line list was compiled and EWs calculated.

 We determined  iron abundances from Y14 EW measurements with the 
package {\tt GALA} \citep{gala} based on the width9 code by Kurucz. 
Model atmospheres were calculated with the {\tt ATLAS9} code
assuming local thermodynamic equilibrium (LTE) and one-dimensional, plane-parallel geometry;
starting from the grid of models available on F. Castelli's website \citep{castelli03}.
For all the models we adopted as input metallicity in the iron abundance derived by Y14.
We adjust the input metallicity to the value we find in the subsequent iterations.
The {\tt ATLAS9} models employed were computed with the new set of opacity 
distribution functions \citep{castelli03} and exclude approximate overshooting in 
calculating the convective flux.

For the abundance analysis, we used only lines with a reduced EW (EWr = $\log$(EW/$\lambda$)
between --5.5 (corresponding to $\simeq$20~m\AA~for a line at 6300\AA) and --4.6 (corresponding to $\simeq$158~m\AA~for a line at 6300\AA), in order to 
avoid lines that are both weak and noisy and to exclude lines in the 
flat part of the curve of growth, respectively.
To compute the abundance of iron, 
we kept only lines within 3$\sigma$ from the median iron value.
The adopted reference solar 
values are from \citet{grevesse98}.

  \begin{figure*}
  \begin{center}
\includegraphics[width=2.3\columnwidth]{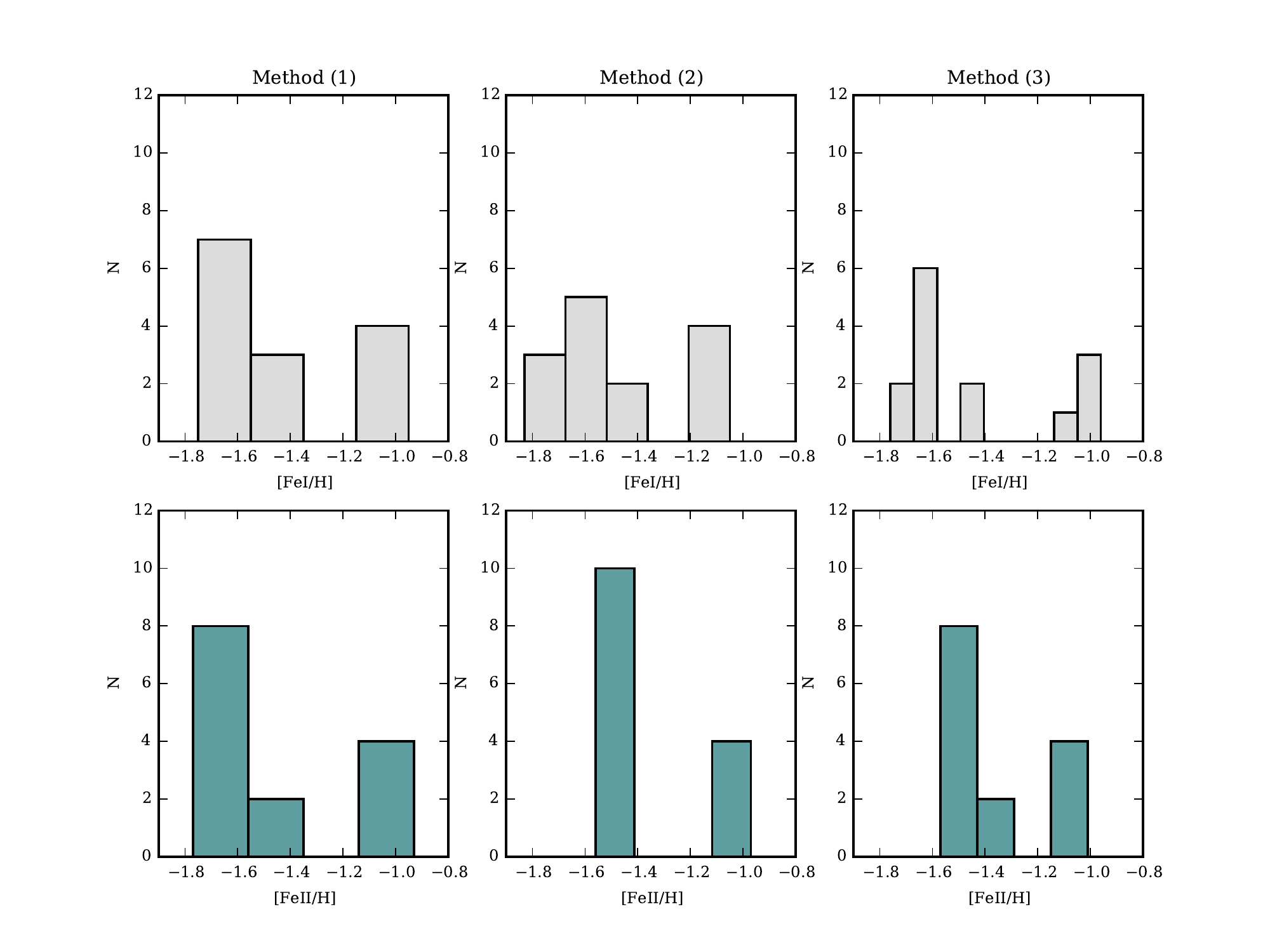}
\caption{ {\em From left to right}: Histograms for [Fe\,{\sc i}/H] (grey) and [Fe\,{\sc ii}/H] (green) 
obtained from the analysis performed with the spectroscopic, photometric, and the {\em hybrid} analysis (method {\em 1},  {\em 2}, and {\em 3} in Section~\ref{gspectro} respectively). {\em Note that histograms represent a severely biased sample and only $\simeq$1-3\% of the cluster stars are metal rich} \citep[e.g.][]{milone15}.}
        \label{FEMIO}
\end{center}
  \end{figure*}

We test how the approach chosen to constrain  the stellar effective temperature and surface gravity  
impacts on the measured  [Fe\,{\sc i}/H]  and  [Fe\,{\sc i}/H] abundance ratios by performing three
independent analysis. 

\begin{enumerate}[label=({\arabic*})]

\item {\em Spectroscopic analysis}. \\
Firstly, we adopt a traditional spectroscopic approach, as done by Y14, in order to verify whether we obtain the 
same evidence of a metallicity dispersion. 

Atmospheric parameters are constrained as follows.
The effective temperature (T$_{\rm{eff}}$) is adjusted until there is no trend between the abundance from 
Fe\,{\sc i} lines and the excitation potential (EP). The surface gravity ($\log$ g) is 
optimised in order to minimise the difference between the abundance derived from neutral and single 
ionised iron. Finally, the microturbulent velocity ($\xi_{\rm{t}}$) is established by erasing 
any trend between the abundance from Fe\,{\sc i} and EWr.  
The same approach to constrain $\xi_{\rm{t}}$
is used consistently in the three independent analysis. 
The atmospheric parameters are then set in an iterative fashion, 
first setting the temperature, and then revisiting T$_{\rm{eff}}$ as the gravity and microturbulence are tweaked. 

The error associated with the determination of $\xi_{\rm{t}}$ 
is estimated by propagating the uncertainty in the slope in the rEW versus Fe abundance plane.
This uncertainty is of the order of $\simeq$0.07 km sec$^{-1}$.
The derived T$_{\rm{eff}}$ are typically based on more than $\simeq$ 70-130 Fe\,{\sc i} lines and have 
internal uncertainties of about 50-60 K, while the internal uncertainties in log g are of the order of $\simeq$0.06 dex.

The derived atmospheric parameters and [Fe\,{\sc i}/H] and [Fe\,{\sc ii}/H] abundances are listed in Table~\ref{FERRO_TAB}, along with 
their associated uncertainties. The left-hand panel of Figure~\ref{IRON} illustrates how they compare to Y14 results.

We observe that our T$_{\rm{eff}}$, log g, and $\xi_{\rm{t}}$ estimates are in excellent agreement with those listed in Y14,
being the average difference between our and Y14 estimates $\Delta$T$_{\rm{eff}}$=--19 $\pm$ 8 K, 
$\Delta$$\log$ g=--0.0 $\pm$ 0.0 dex, $\Delta$ $\xi_{\rm{t}}$=--0.09 $\pm$ 0.02 km sec$^{-1}$.
The agreement between our [Fe\,{\sc i}/H]  and [Fe\,{\sc ii}/H]  abundances and those listed in Y14 is also very good.
In particular, we note that the metal-poor stars identified by Y14, i.e. see Table~\ref{PHOT_PAR}, are also the most metal-poor stars in 
our analysis (see Table~\ref{FERRO_TAB}). Additionally, the four stars with [Fe/H]$\simeq$--1.0 dex 
are the same metal-rich stars in Y14.

In the left-hand panel of Figure~\ref{FEMIO}, we plotted the histograms of the 
[Fe\,{\sc i}/H]  and [Fe\,{\sc ii}/H] distributions when the full spectroscopic analysis is performed.
Our ability to portrait the [Fe\,{\sc i}/H]  and [Fe\,{\sc ii}/H] distributions for this small dataset 
critically depends on the adopted bin size ($\Delta$). To adequately represent the 
shape of the underlying distribution, in Figure~\ref{FEMIO} we selected $\Delta$  as 
the value which minimise $C(\Delta)=(2 me -va)/\Delta^{2}$; where $me$ and $va$ are the mean and variance of the dataset.

As one can see, we obtain an iron dispersion that mirrors the dispersion measured by Y14. 
We find a large ($\Delta$[Fe/H] $\simeq$ 0.8 dex) iron distribution with two prominent peaks  at
[Fe/H] $\simeq$--1.7 and --1.0 dex and a smaller component at [Fe/H] $\simeq$--1.5 dex.
The iron dispersion within the metal-poor stars is larger ($\Delta$[Fe/H] $\simeq$ 0.4 dex) than that expected from observational 
errors alone, indicating the presence of two separated subpopulations with different metallicity.\\

\item {\em Photometric analysis}. \\
We use $VI$ WFI photometry to derive  photometric T$_{\rm{eff}}$ and $\log g_{*}$ estimates\footnote{The main result of this paper remains unchanged for different choice of colours, as shown in Figure~4 by Mu15.}.
Input T$_{\rm{eff}}$ values have been computed by means of the ($V-I$)$_{0}$-T$_{\rm{eff}}$ transformation by \citet{alonso99}
adopting a colour excess $E(B - V)$ = 0.02 mag taken from the 
most update version (2010) of the McMaster catalog \citep{harris96}.
Surface gravities have been computed assuming the photometric 
T$_{\rm{eff}}$, the apparent visual distance modulus of 15.50 \citep{harris96}  and an evolutionary mass of 
0.82 M$_{\sun}$ \citep{berg01}. Assuming that all the stars in M~2 have the same age\footnote{According to
\citet{milone15} the populations at [Fe/H]$\simeq$--1.7 and --1.0 dex are coeval within $\simeq$ 1 Gyr.}, the 
estimated mass for metal-rich stars is 0.85 M$_{\sun}$. Differences in masses of the order of a few percent of 
M$_{\sun}$ have null impact on the metallicity determination (on the order of 0.01 dex or less).
The bolometric corrections are calculated according to \citet{alonso99}. 

Uncertainties in T$_{\rm{eff}}$ are estimated by taking into account the uncertainty in $V$ and $I$ 
magnitudes and in the colour excess. 
The total uncertainty corresponds to a 
typical uncertainty in T$_{\rm{eff}}$  of $\simeq$ 80-90 K.
As discussed in Section~\ref{OSSERVAZIONI}, the star NR~37 appears to lie slightly off the main RGB in our WFI photometry, while 
it is well located on the main RGB body in the HST photometry by \citet{sarajedini07}. The observed difference in both $V$ and $I$ magnitudes from 
are $\Delta V_{{\rm WFI-HST}}$=--0.085 and $\Delta I_{{\rm WFI-HST}}$=--0.051; and lead to a difference of $\simeq$ 50K in the estimate  T$_{\rm{eff}}$ which is within the errors quoted in Table~\ref{FERRO_TAB}.

Uncertainties in the surface gravity are derived by considering the error sources in T$_{\rm{eff}}$, 
luminosity, and mass. Summing all these terms in quadrature leads to a typical uncertainty in $\log  g_{*}$ of $\simeq$ 0.08 dex.
In the following we assume the (very conservative) error of 0.1 dex in $\log  g_{*}$ for all the targets.
\citet{milone15} found that stars with [Fe/H] = --1.7 dex and very enhanced Na and Al abundances in Y14  
are also enhanced in helium up to Y = 0.315 ($\simeq$17\% of M~2 stars). According to a set of BASTI isochrones with He-enhancement \citep{pietrinferni04,pietrinferni06}, for a metallicity [Fe/H]=--1.5 dex and T$_{\rm{eff}}$=4500 K, stars with Y = 0.315 have surface gravity $\simeq$--0.10 dex lower than that derived for stellar models without He-enhancement. Variations up to $\simeq$--0.10 dex  in   $\log  g_{*}$ have negligible impact (of the order $\simeq$ 0.03 dex) on metallicity determinations from  Fe\,{\sc ii}  lines.

The middle panel of Figure~\ref{IRON} illustrates how (from top to bottom) the measured  [Fe\,{\sc i}/H], and [Fe\,{\sc ii}/H]  abundances, T$_{\rm{eff}}$, $\log  g_{*}$, and  $\xi_{\rm{t}}$ compare to the Y14 results.
 Again, the agreement between the atmospheric parameters obtained as detailed above and Y14 parameters is very good.
 On average, we determine  slightly lower  T$_{\rm{eff}}$ and $\xi_{\rm{t}}$ than Y14, i.e. $\Delta$T$_{\rm{eff}}$=--58 $\pm$ 15 K and 
 $\Delta$ $\xi_{\rm{t}}$=--0.11 $\pm$ 0.01 km sec$^{-1}$ respectively, while the photometric gravities 
 are on average larger than Y14 ones by $\Delta$$\log$ g=0.2 $\pm$ 0.1 dex ($\sigma$=0.2 dex).

The middle panel of Figure~\ref{IRON} also shows the difference between neutral and single ionised Fe abundances and Y14 
measurements. The  Fe\,{\sc i} and Fe\,{\sc ii} distributions are clearly different. 
The iron distribution obtained from Fe\,{\sc i} lines resembles that obtained with method {\em (1)}, with a metal rich component at 
[Fe\,{\sc i}/H] $\simeq$--1.0 dex and a metal-poor group with iron abundances in the range between  [Fe\,{\sc i}/H] = --1.83 to --1.40 dex. 
The dispersion of [Fe\,{\sc i}/H]  abundances in the metal-poor stars is larger than those expected from measurement errors alone, 
indicating the presence of an {\em intrinsic} spread. Conversely, the distribution obtained from Fe\,{\sc ii}  lines is bimodal, with a 
metal poor ([Fe\,{\sc ii} /H] $\simeq$ --1.5 dex)  and a metal-rich ([Fe\,{\sc ii} /H] $\simeq$ --1.0 dex) components.

 The [Fe\,{\sc i}/H] and [Fe\,{\sc ii}/H] abundances obtained with this method are listed in Table~\ref{FERRO_TAB}.
 To better visualise results, we present histograms for the Fe\,{\sc i} and Fe\,{\sc ii} distributions in the middle panel of 
 Figure~\ref{FEMIO}. The Fe\,{\sc i} distribution shows again two main metallicity populations,  stars with --1.8 $\leq$ [Fe/H] $\leq$ --1.4  dex showing 
 also an internal iron dispersion.
 On the other hand, the Fe\,{\sc ii} distribution is clearly bimodal. Additionally, the dispersion within each group is consistent with the uncertainties, suggesting the presence of only two population, neither with any intrinsic dispersion.

\item {\em Hybrid analysis}. \\
Finally, we repeat the analysis adopting spectroscopic temperatures, 
i.e. T$_{\rm{eff}}$ is adjusted until there is no slope between the abundance from 
Fe\,{\sc i} lines and the EP. The advantage of this {\em hybrid} analysis is twofold.
Firstly, the large number of Fe\,{\sc i} lines distributed over a large range of EPs, allows for a 
very accurate spectroscopic T$_{\rm{eff}}$, with internal uncertainties as small as in the spectroscopic method.
Secondly, the T$_{\rm{eff}}$ estimated in this fashion are not greatly affected by the uncertainties in the photometry, and in the differential and absolute reddening which impact on the  photometric T$_{\rm{eff}}$ determination.

The gravity $\log  g_{*}$  is computed through the Stefan-Boltzmann equation 
according to the new spectroscopic  value of T$_{\rm{eff}}$ at each iteration.  We adopt the same values for 
the distance modulus, the stellar mass and bolometric corrections as in the photometric analysis.

Table~\ref{fe2_EW} lists all the Fe\,{\sc ii} lines used or discarded in our analysis with respect to \citet{yong14} study. Again, atomic data and EWs are from \citet{yong14}.
Generally, we adopted a more stringent rejection for very weak lines (of the order of $\simeq$15 m\AA), however, the number of lines considered to infer abundances is comparable with \citet{yong14}.

The T$_{\rm{eff}}$ and $\xi_{\rm{t}}$ measured in this fashion are again in excellent agreement with Y14, as
shown in the right-hand panel of Figure~\ref{IRON}. As in the case of method {\em (2)}, the photometric gravities are on average 
larger, i.e. $\Delta$ $\log$ g=0.2 dex, than the spectroscopic ones. 
As for method {\em (2)}, the new choice of atmospheric parameters does not wipe out the difference between the [Fe\,{\sc i} /H] and [Fe\,{\sc ii} /H] distributions. The iron abundances measured from  [Fe\,{\sc i} /H] lines remain very different than those obtained from  [Fe\,{\sc ii} /H]  lines (see Table~\ref{FERRO_TAB}).
This is also evident from the right-hand panel of Figure~\ref{FEMIO}, where the neutral 
and single ionised iron distributions are shown as histograms.

\end{enumerate}

\subsection{Uncertainty determinations}
Uncertainties on the derived abundances have been computed for each target by adding in quadrature the two main error sources: 

\begin{enumerate}
\item uncertainties arising from the EW measurements, which have been estimated as the line-to-line abundance scatter divided by the square root of the number of lines used. This term is of the order of 0.01-0.02 dex for Fe\,{\sc i} and 0.03-0.05 dex for Fe\,{\sc ii}.
\item uncertainties arising from the atmospheric parameters.
	Those are computed varying by the corresponding uncertainty only one parameter at a time, while keeping the others fixed in the 
	photometric analysis, i.e., case {\em (2)}. In this case the total uncertainties in [Fe\,{\sc i}/H] are of the order of 0.07-0.08 dex, while in [Fe\,{\sc ii}/H] are of about 0.10-0.11 dex  (due to the higher sensitivity of Fe\,{\sc ii}  lines to T$_{\rm{eff}}$ and $\log$ g). 
	
	When the temperature is spectroscopically optimised, i.e, cases {\em (1)} and {\em (3)},  the uncertainties have been computed following the approach described by \citet{cayrel04}  to take into account the covariance terms due to the correlations among the atmospheric parameters. For each target, the temperature has been varied by $\pm$ 1 $\sigma$ $_{T_{\rm{eff}}}$ , the gravity has been recomputed using the Stefan-Boltzmann equation adopting the new values of T$_{\rm{eff}}$ and the  v$_{t}$  derived spectroscopically. The total uncertainties  in [Fe\,{\sc i}/H] are of the order of 0.05 dex, while in [Fe\,{\sc ii}/H] are of  about 0.07-0.08 dex.
\end{enumerate}

Table~\ref{FERRO_TAB} lists both the  uncertainty originating from EW measurements ($\delta_{\rm{int}}$), the uncertainty due to atmospheric parameters  ($\delta_{\rm{par}}$). 

  \begin{figure}
 \centering
\includegraphics[width=0.9\columnwidth]{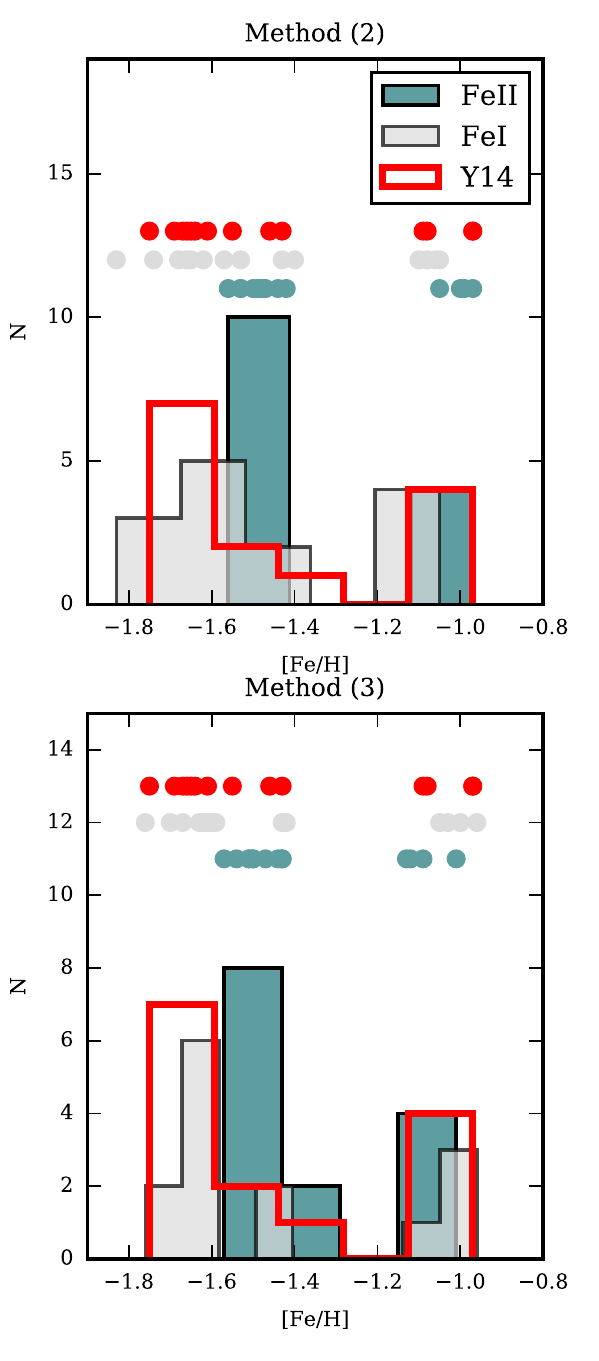}
\caption{Distribution of  Fe\,{\sc i} (grey histograms) and  Fe\,{\sc ii} (green histograms) metallicity from our abundance analysis 
using the {photometric} and {\em hybrid} approach of Section~\ref{gspectro} (top and bottom panels, respectively).
For comparison, the metallicity distribution as measured by \citet{yong14} is shown in red.
The actual data points used to construct the histogram are shown above the graph.}
        \label{YONGFE}
   \end{figure}
  
 \section{Discussion}\label{DISCUSSION} 
 The spectroscopic analysis of Section~\ref{gspectro} confirms Y14 results:  when analysed with atmospheric parameters derived following the traditional spectroscopic approach, the stars with [Fe/H] $\leq$--1.5 dex reveal a clear star-to-star scatter in the iron content. 
The metallicity distribution from Fe\,{\sc i}  lines ( grey histograms in Figure~\ref{FEMIO}) is still large with three obvious metallicity sub-populations when photometric gravities are adopted --i.e. methods {\em (2)} and {\em (3)}, while the distribution derived from Fe\,{\sc ii} lines is clearly bimodal. The two metallicity components at [Fe\,{\sc ii} /H]=--1.5 and --1.1 dex do not show any internal iron spread. 

It must be noted that Fe distributions of Figure~\ref{FEMIO} are not actually representative of the cluster iron distribution.
The spectroscopic sample includes four metal-rich and ten metal-poor stars. Metal-rich stars account only for 
$\sim$1-3\% of M~2 stars \citep{milone15}, therefore the metal-rich RGB is better sampled than the main RGB in Y14 data.
{\em As a result, a Fe\,{\sc ii} distribution, weighted based on the fraction of stars observed in the main and metal-rich RGB,
would display a large, unimodal component at  [Fe\,{\sc ii}/H]$\simeq$--1.5 dex and a negligible component ($\leq$3\%) at [Fe\,{\sc ii}/H]$\simeq$--1.1 dex.}

\begin{figure}
 \begin{center}
\includegraphics[width=0.9\columnwidth]{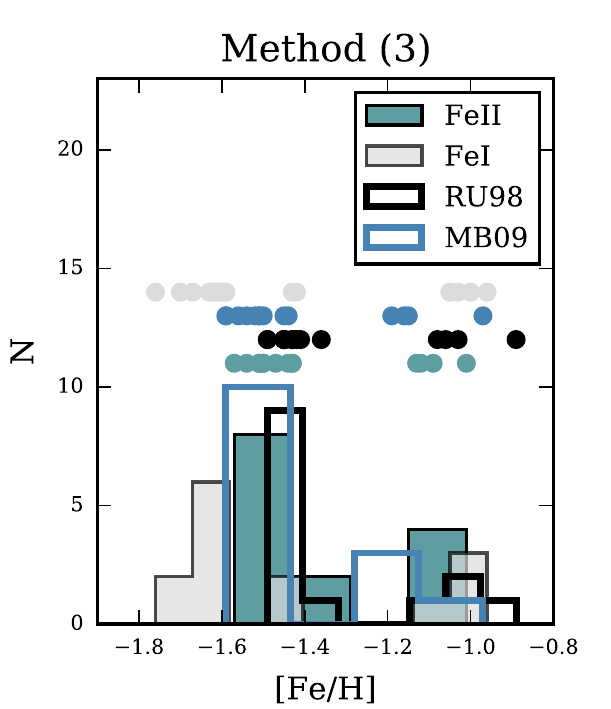}
\caption{Distribution of [Fe\,{\sc i} /H] (grey histogram) and [Fe\,{\sc ii} /H]  (green histogram) abundance ratios, when using the {\em hybrid} analysis of Section~\ref{gspectro}. The blue and black empty histograms are the distributions for Fe\,{\sc ii} lines obtained using different literature sources for 
$\log g_{f}$ values;~\citet{melendez09} and \citet{raassen98} respectively. Note that the Fe\,{\sc ii}  remains bimodal irrespective on the 
choice of atomic data. In particular, Fe\,{\sc ii}  distributions lack the metal-poor tail seen in Fe\,{\sc i} histogram.}
     \label{LOGG}
     \end{center}
  \end{figure}

In Figure~\ref{YONGFE}, we plot the Fe\,{\sc i}  and Fe\,{\sc ii} distributions presented by Y14 and the iron distributions we obtained using the {\em photometric} and
 {\em hybrid} approach of Section~\ref{gspectro}. Grey and green symbols
are individual measurements for [Fe\,{\sc i}/H] and   [Fe\,{\sc ii}/H], respectively; while red symbols are Y14 [Fe\,{\sc i}/H] measurements\footnote{Since Y14 adopted the classic spectroscopic approach, Fe\,{\sc i} abundances are set to be equal to Fe\,{\sc ii} ones {\em by construction.}}. Figure~\ref{YONGFE} shows how the range in metallicity is 
largely diminished when adopting  Fe\,{\sc ii} abundances and and photometric gravities.
This remains true also when considering different $\log g_{f}$ for the Fe\,{\sc ii} (e.g., \citealp{melendez09,raassen98}; see Figure~\ref{LOGG}).
This demonstrates that the main RGB of M~2 has a homogeneous metallicity of [Fe/H] $\simeq$ --1.5 dex, with
 a very small spread which is comparable to the star-to-star variations observed in other GCs. On the other hand we confirm the presence of a second metallicity sub-population at higher metallicities.
 
 In the following, we restrict ourselves to the metal-poor component, i.e. 
 stars with  [Fe\,{\sc ii}/H] $\simeq$--1.5 dex.
According to our {hybrid analysis}, stars NR~76 and NR~77 are both RGB stars with similar atmospheric parameters,
i.e. T$_{\rm{eff}}$/$\log$ g/ v$_{t}$=4347-4339K/1.07-1.10 dex/1.6-2.10 kms$^{-1}$, respectively.
Nonetheless, they display very different [Fe\,{\sc i}/H] -- [Fe\,{\sc ii}/H] difference. Star NR~76 has [Fe\,{\sc i}/H] -- [Fe\,{\sc ii}/H]= --0.16 dex,  while for 
star NR~77  [Fe\,{\sc i}/H] and [Fe\,{\sc ii}/H]  lines return basically the same Fe abundance. 

NLTE effects lead to a systematic underestimate of the EWs from neutral lines, i.e. Fe\,{\sc i} lines. Therefore, the standard LTE analysis of lines formed in NLTE gives lower abundance from neutral lines \citep{mashonkina11}. 
Although the available non local thermodynamic equilibrium (NLTE) calculations do not foresee different corrections for stars with similar atmospheric parameters  \citep[e.g.][]{bergemann12}, the observed discrepancy between abundances inferred by Fe\,{\sc i} and Fe\,{\sc ii} lines can be qualitatively explained by the occurrence of some effects driven by overionisation, i.e. affecting mainly the less abundant species (Fe\,{\sc i}; \citealp{fabrizio12}). 

We find that spectroscopic gravities are, on average, lower than the photometric ones by $\simeq$0.2 dex, with a maximum difference of $\simeq$0.6 dex
As an example, the {\em hybrid} analysis of star NR~60 provides [Fe\,{\sc i}/H] = --1.76 $\pm$ 0.01 dex and  [Fe\,{\sc ii}/H] =--1.51 $\pm$  0.03 dex, with 
T$_{\rm{eff}}$=4318 K, $\log$ g=0.92 dex, and v$_{t}$ =2.1 kms$^{-1}$. When a fully spectroscopic analysis is performed 
we obtain T$_{\rm{eff}}$=4318 K, $\log$ g=0.30 dex, and v$_{t}$ = 2.1 kms$^{-1}$ and the measured abundances are 
[Fe\,{\sc i}/H] = -- 1.75 $\pm$ 0.01 dex and  [Fe\,{\sc ii}/H] =--1.77 $\pm$  0.03 dex. Thus, the spectroscopic values of 
 T$_{\rm{eff}}$ and v$_{t}$ are virtually the same to those adopted in the {\em hybrid} approach. Hence, 
 the [Fe\,{\sc i}/H] and [Fe\,{\sc ii}/H]  distributions  are only marginally affected by the choice of different methods 
 to optimise T$_{\rm{eff}}$ (see for example Figure~\ref{IRON}).  
 
 \begin{figure}
 \begin{center}
\includegraphics[width=0.85\columnwidth]{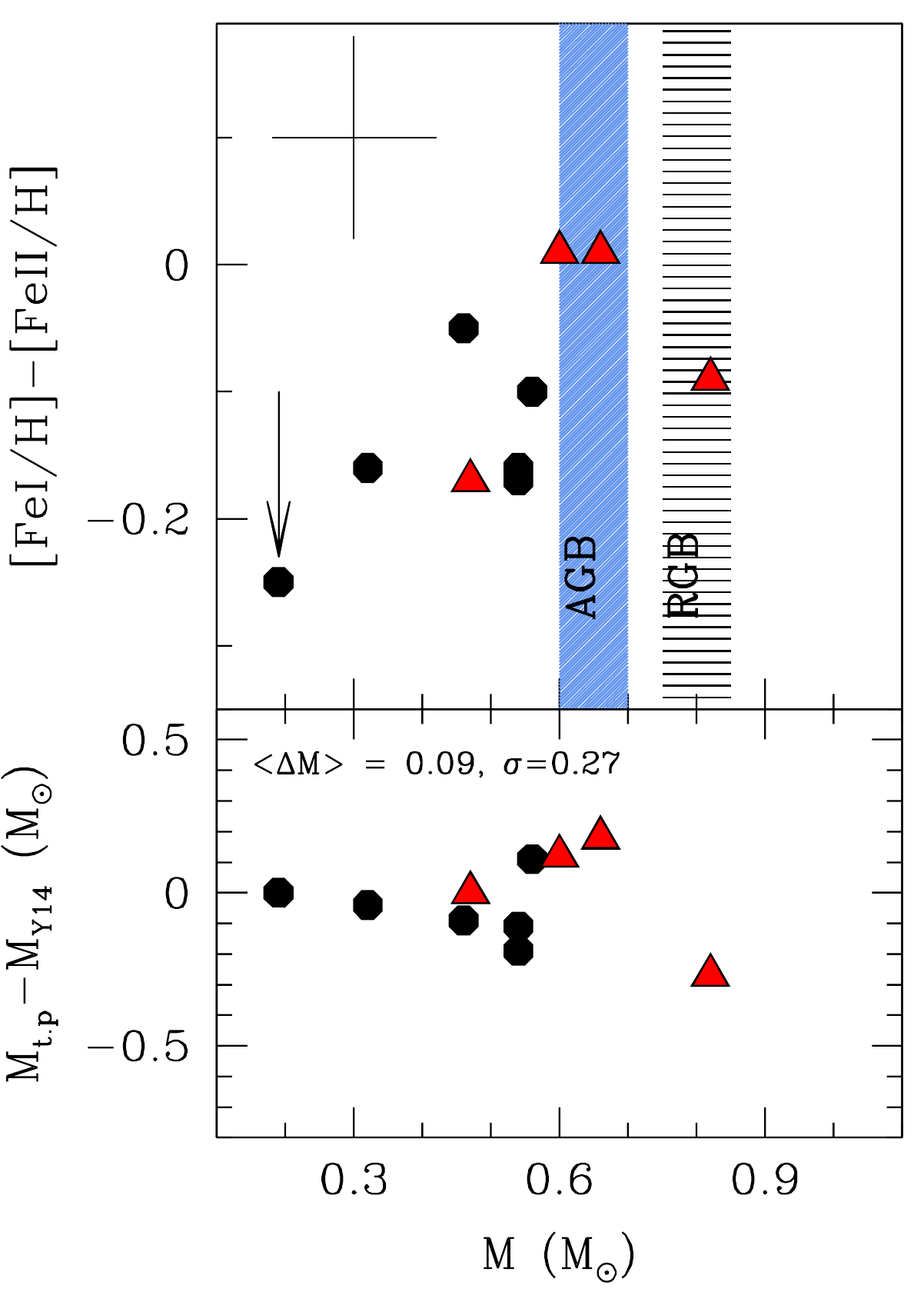}
\caption{{\em Top panel:} the difference  [Fe\,{\sc i} /H]--[Fe\,{\sc ii} /H] (as derived using the {\em hybrid} analysis of Section~\ref{gspectro}) 
is plotted against the stellar masses 
inferred from Y14 analysis. The shaded and blue regions define the mass range expected for RGB and AGB stars, respectively.The arrow indicates NR~60, a likely AGB star
according to its position in the CMD (see Section~\ref{OSSERVAZIONI}).
The typical error is also plotted in the right upper corner.
Symbols are as in Figure~\ref{CMD}. {\em Bottom panel:} we plot the difference in mass we obtain when stellar masses are 
estimated using our {\em spectroscopic} and Y14 parameters; M$_{{\rm t.p}}$  and M$_{{\rm Y14}}$ respectively.}
     \label{MASS}
     \end{center}
  \end{figure}

\subsection{Stellar masses inferred from spectroscopic gravities}\label{mass-grav}
While the [Fe\,{\sc i}/H] and [Fe\,{\sc ii}/H]  distributions  are only to some degree changed by the choice of different methods 
 to constrain T$_{\rm{eff}}$, the choice of spectroscopic gravities can heavily affect the measured abundances and hence the conclusions 
 about the degree of homogeneity of the cluster (see Figure~\ref{IRON}).
 
Different methods to derived $\log g$, i.e. photometry, ionisation balance or pressure-broadened wings of strong lines, can provide conflicting results (see e.g. \citealp{edvardsson88,gratton1996, fuhrmann98, allende99}).

A simple but sound method to check the validity of the spectroscopic gravities is to derive the stellar masses that these gravities imply. In the case of globular cluster stars, where ages and distances are known and are the same for all the stars in a given clusters, the RGB stars will cover a small range of masses and they are expected to have masses larger than 0.5 M$_{\odot}$, corresponding to the mass of the He-core for low-mass stars.
 We estimate stellar masses as inferred from spectroscopic gravities for stars with [Fe/H]$\simeq$--1.5 dex; i.e. the 
metal-poor and metal-intermediate components in Y14 analysis.
We compute stellar masses from the Stefan-Boltzmann equation, starting from the surface gravity values constrained with method {\em (1)} in Section~\ref{gspectro}. We note that the results do not change when using the spectroscopic gravities listed in Y14, 
being the average difference between our and Y14 mass estimates $\Delta$ M=0.09 M$_{\sun}$ (see bottom panel in Figure~\ref{MASS}, where 
the inferred stellar masses are compared to each other).

The estimated masses are plotted against the difference in [Fe\,{\sc i}/H]--[Fe\,{\sc ii}/H] from method {\em (3)} in the top panel of Figure~\ref{MASS}. 
Firstly, we note that stars for which  [Fe\,{\sc i}/H] and  [Fe\,{\sc ii}/H] lines give the same iron abundance,
i.e. stars with [Fe\,{\sc i}/H]--[Fe\,{\sc ii}/H]$\simeq$ 0, have an estimated mass compatible to that expected for a giant star in a GC
(M$\simeq$0.75-0.85 M$_{\sun}$)\footnote{We stress, however that different choices of distance modulus or different $V$ magnitudes for target stars 
can cause shifts in the mass distribution, but they do not affect the overall mass distribution.}. Those stars --NR~47 and NR~77-- are also the stars for which 
the ionisation balance gives the same surface gravities along the mean RGB as photometry, and they are also the most metal-rich stars
of the canonical RGB according to Y14 analysis
(with [Fe/H]=--1.42 and --1.46 dex). 

Figure~\ref{MASS} also illustrates how the estimated spectroscopic stellar masses show a very large spread which is not expected for 
stars in the same evolutionary stage. Indeed, the derived masses range from 0.19 M\sun~for NR 60 (a very unlikely value for a giant even taking in to account the uncertainties in the mass loss rate) to 0.82 M\sun~for NR 81. Stars with very (unphysical) low masses display the largest difference in the [Fe\,{\sc i}/H]--[Fe\,{\sc ii}/H]. On the contrary, when the [Fe\,{\sc i}/H] and [Fe\,{\sc ii}/H] abundance ratios are similar, the spectroscopic gravities provide stellar masses that are compatible within the theoretical expectation (Figure~\ref{MASS}). 


Interestingly,  NR~60 --the star showing the largest [Fe\,{\sc i} /H]--[Fe\,{\sc ii} /H] difference when 
photometric gravities are adopted, has been classified as an AGB star by Y14 and its position appears to be consistent with that of an AGB also in our own photometry (see Section~\ref{OSSERVAZIONI}).
The discrepancy between iron abundances measured by Fe\,{\sc i} and Fe\,{\sc ii} lines for AGB GC stars
has been extensively reported in literature in the last few years. Indeed, Fe\,{\sc i} lines have been found to provide 
 systematically lower abundances in AGBs with respect to RGBs of the same cluster. This behaviour has been observed in M5 \citep{ivans01}, 47~Tuc \citep{lapenna14}, NGC~3201 \citep{mucciarelli15}, M22 (Mu15) and M62 \citep{lapenna15}.


As discussed before, Y14 find that metal-intermediate stars in their analysis are also enriched in their $s$-process
element content with respect to metal-poor stars. Unfortunately, $s$-element abundances have been derived from spectral synthesis by Y14 and only spectra for the nine stars observed with SUBARU are publicly available (3 metal-poor and 2 metal-intermediate stars). 
Therefore, we do not perform any attempt to measure $s$-element abundances. However, we can tentatively assume 
that $s$-rich stars are still enriched in their $s$-process element content also when [Fe\,{\sc ii}/H] abundances from
 photometric gravities are taken as a reference.

It is conceivable that this can be the case, as in \citet{lardo13} we demonstrate that the split RGB of M~2 observed in the 
$U$, $U-V$ CMD \citep{lardoM2}, is composed by two groups of stars which differ by $\simeq$0.5 dex in their $s$-process element 
content\footnote{We also note that intermediate-metallicity stars are enriched by a similar amount in their [Y\,{\sc ii}/Fe] abundances,
i.e. $\Delta$ [Y\,{\sc ii}/Fe] $\simeq$0.5 dex , according to Y14 (see their Figure~11).}.
The analysis presented in \citet{lardo13} is based on low-resolution spectra, and [Sr\,{\sc ii}/Fe] and [Ba\,{\sc ii}/Fe] abundances were measured assuming the same metallicity for all stars.
Therefore, the observed spread in the absolute abundances for  [Ba\,{\sc ii}/Fe] and  [Sr\,{\sc ii}/Fe] is larger than that measured from 
[Fe\,{\sc i}/H] lines ($\simeq$0.2 dex), possibly indicating that $s$-rich stars in Y14 analysis remain $s$-rich also when 
[Fe\,{\sc ii}/H] abundances derived from photometry are used.

Under this assumption, Figure~\ref{MASS} illustrate that the difference between [Fe\,{\sc i}/H]--[Fe\,{\sc ii}/H] is also correlated with the $s$-process element content, having $s$-rich stars better agreement between [Fe\,{\sc i}/H]--[Fe\,{\sc ii}/H]  than $s$-normal stars (red triangles and black dots, respectively). In these respects, the main metal-poor
 component of M~2 is very similar to M~22, where a correlation between the difference between [Fe\,{\sc i}/H]--[Fe\,{\sc ii}/H]  and the $s$-process element content also exists (Mu15).

 \section{Summary and Conclusions}\label{CONCLUSIONI}
 
In this paper we re-derive iron abundances from EW measurements of 14 M~2 RGB stars from which 
Y14 detect a large metallicity dispersion ($\simeq$ 0.8 dex) with three main components
at [Fe/H]$\simeq$--1.7, --1.5, and --1.0 dex.
[Fe\,{\sc i}/H] and [Fe\,{\sc ii}/H] abundances have been calculated using three different approaches to constrain 
effective temperature and surface gravity. 
We can summarise our main results as follows:

\begin{itemize}

\item We find that the [Fe\,{\sc i}/H] and [Fe\,{\sc ii}/H]  distributions greatly
differ when spectroscopic or photometric gravities are adopted in the analysis.
In particular, the distribution of [Fe\,{\sc i}/H]  abundances remains large (and possibly trimodal) 
irrespective of the choice of spectroscopic or photometric gravities. 
On the other hand, when photometric gravities are used, the [Fe\,{\sc ii}/H] distribution is clearly bimodal,
with two separate components at [Fe/H]$\simeq$--1.5 and --1.1 dex.
Both stellar groups are compatible with no internal iron spread.

{\em  Hence, the large majority of cluster's stars, i.e. $\simeq$ 99\% of the total cluster population, do not show any evidence for a metallicity spread.} Stars in this metallicity group show the C-N, Na-O anti-correlations typical of GCs, as well as a bimodality
in their $s$-process content (\citealp{lardoM2,lardo13}, Y14) which is somewhat connected to  
the observed discrepancy between Fe\,{\sc i} and Fe\,{\sc ii} abundances (see Setion~\ref{mass-grav}).
On the contrary, we confirm the presence of a second metallicity component \citep{yong14}.
Stars at [Fe\,{\sc ii}/H] $\simeq$ --1.1 dex, if members, 
represent a very small fraction of M~2 stars, accounting only for $\sim$1\% of the  cluster population \citep{milone15}.

\item The results presented in this paper for the metal-poor stars in M~2 are similar to what found by Mu15 in M~22.
The absence of a metallicity spread among M~22 stars claimed by Mu15 (but see \citealp{dacosta09}, and \citealp{marino11}) would rule out the hypothesis that the cluster was significantly more massive in the past and possibly the remnant of a now disrupted dwarf galaxy \citep[see][for a discussion on the mass loss problem in GCs]{bastian15}. 
We stress, however, that our study confirms the presence of 
a second minority population with different iron abundance with respect to the bulk of M~2 stars \citep{yong14}.
This additional metal-rich population appears not to be 
present in M~22 (Mu15;  but see \citealp{dacosta09}, and \citealp{marino11}).

\item Fe\,{\sc i} lines are commonly used to derive iron abundances, because
 a considerable number of Fe\,{\sc i} lines in a rather large range of EP  with accurate transition probabilities exists \citep[e.g.][]{blackwell79b}.
However, even if accurate laboratory atomic data are available for a limited number of Fe\,{\sc ii}  lines \citep[see Appendix A2 of][for an extensive investigation of the reliability of $\log g_{f}$ values for Fe\,{\sc ii}]{lambert96}; in the atmospheres of giant stars iron is almost completely ionised \citep{kraft03} and Fe\,{\sc ii} lines do not suffer for departure from LTE, at variance with Fe\,{\sc i} ones. As a result, metallicity is more safely derived from the dominant species, Fe\,{\sc ii}, than from Fe\,{\sc i}.

We consistently used the same Fe\,{\sc ii}  data for all the programme stars. Therefore, the uncertainty in the Fe\,{\sc ii} $\log g_{f}$ values is expected to touch only  the zero-point (i.e. the {\em absolute} value) of the [Fe\,{\sc ii}/H] abundances, while leaving  
 the [Fe\,{\sc ii}/H] distribution of the entire stellar sample unaffected. Indeed, the [Fe\,{\sc ii}/H] distribution is bimodal even when considering different sources for the atomic data (e.g., \citealp{melendez09, raassen98}; see Figure~\ref{LOGG}).

 The same number of Fe\,{\sc ii}  lines ($\simeq$10-20) used in this paper to infer metallicities has been used by Y14  to derive atmospheric parameters; and the atmospheric parameters derived in such fashion would not be considered as unreliable due to the small number of lines of Fe\,{\sc ii} used to constrain gravity.
 
\item From a qualitative point of view, the (negative) difference between [Fe\,{\sc i}/H] and [Fe\,{\sc ii} /H] observed in some M~2 stars is consistent with the occurrence of NLTE effects driven by overionisation, as the most metal-poor stars in the [{Fe\,{\sc i}/H] distribution are shifted to higher metallicities in the [Fe\,{\sc ii}/H] distribution.  The same holds also for different choices of literature sources for the $\log g_{f}$ (e.g., \citealp{melendez09, raassen98}).
  
In the atmospheres of metal-poor giants}, Fe\,{\sc ii} is the dominant species 
 throughout the atmosphere, so that NLTE effects are negligible for Fe\,{\sc ii}, while they are extremely important for Fe\,{\sc i}.
For example, \citet{thevenin99} find that the reduction of [Fe/H] estimated from Fe\,{\sc i} relative to Fe\,{\sc ii} amounts
to about 0.1 dex at [Fe/H] = --1 dex and about  0.3 dex at [Fe/H] = --2.5 dex \citep[see also][]{lambert96}.

However, the observed differences are not quantitative compatible with the theoretical predictions (e.g. \citealp{bergemann12}), suggesting that this interpretation is no correct or that the available NLTE correction grids are not suitable for these stars.


\end {itemize}

The observational evidence collected in the last years provide a more complex picture for the so-called {\em anomalous} clusters, not easy to interpret from a theoretical point of view. 
Observationally, we note that there are two circumstances where the use of spectroscopic gravities leads to spurious iron spreads:

\begin{enumerate}

\item In {\em normal} clusters, i.e. clusters showing {\em only} the characteristic Na-O anticorrelation, the detection of a spurious iron spread 
can be ascribed to the inclusion of AGB stars in the spectroscopic sample, i.e. \citet{mucciarelli15}. When AGB stars are removed from the analysis, such clusters show no evidence for a significant iron spread. This remains true when both photometric and spectroscopic gravities are adopted.
Therefore, it appears not so surprising that {\em normal} clusters result to be mono-metallic also when analysed with spectroscopic gravities, if the sample include only RGB stars. For example, Mu15  compared M~22 stellar metallicities with those of NGC~6752, using the same analysis method. They found no evidence for a significant iron spread among NGC 6752 stars also when atmospheric parameters are derived in a classical spectroscopic fashion. Finally, we note that the largest study surveying GC stars, i.e.  {\em the Na-O anticorrelation and HB} project (see e.g. \citealp{carretta2009,car09b}), carefully avoids to target AGB stars.

\item In {\em anomalous} clusters; i.e. clusters with spreads in $s$-process elements (and possibly C+N+O), double SGB and split RGB (as in the case of M22 and the the majority of M2, see Introduction) a spurious spread is derived when spectroscopic gravities are adopted. Such clusters are indeed mono-metallic when  Fe\,{\sc ii}  lines and photometric gravities are used in the analysis. Again, the group publishing the largest metallicity database for GC stars, i.e. 
the {\em Na-O anticorrelation and HB} project, consistently use photometric data to constrain both effective temperatures and surface gravities. This explains why clusters  showing intrinsic [Fe/H] are relatively rare. For example, \citet{yongM62} performed the same analysis as on M~2 stars in M~62, finding no evidence for iron dispersion. This cluster is the ninth most luminous cluster in the Galaxy and it is characterised by an extended horizontal branch\footnote{The observational evidence indicates that GCs with intrinsic metallicity spreads are  are preferentially the more luminous clusters with extended horizontal branches.}, yet it does show neither a $s$-process element bimodality \citep{yongM62} nor a photometric SGB split. Similarly,  \citet{carretta362} did not report a metallicity spread for stars in NGC~362, an {\em anomalous} cluster with mass comparable to that of M~22, which shows a split SGB, a multimodal RGB, and a spread in its $s$-process element content. To constrain atmospheric parameters, \citet{carretta362} followed the same procedure adopted for the other GCs targeted by their FLAMES survey, i.e. photometric stellar gravities.

All the above is consistent with our hypothesis that spectroscopic gravities yield artificial iron spreads only in {\em anomalous} clusters, a relatively small subset of the Milky Way GC population.

\end{enumerate}

So far, a number of GCs have been claimed to host different populations of stars with different metallicities.
Large iron distributions have been found in $\omega$~Centauri  \citep[e.g.]{jhonson10}, M~54 \citep{carrettaM54}, and Terzan~5  \citep[e.g.][]{massari14}. Additionally, smaller intrinsic spreads, i.e. 
comparable to those found by \citet{marino09} and Y14 in the case of M~22 and the main metal-poor component in M~2, 
are  found in a growing number of GCs.
The analysis presented in Mu15 and in this paper casts doubt upon the presence of these intrinsic spreads, which can 
have been artificially produced by the choices to constrain atmospheric parameters. 
In particular, the use of spectroscopic gravities in both M~22 (Mu15) and M~2 leads to very low stellar masses which are unphysical for a giant star.
Our  findings suggest caution when analysing metal-poor stars using gravities obtained from the ionisation balance, 
and indicates the need to carefully reanalyse clusters found to display an intrinsic iron spread (e.g. M~19, NGC~5286).

\section{acknowledgements}
We thank E.~Pancino, M.~Bellazzini, G.~Altavilla, G.~Cocozza, S.~Marinoni, S. Ragainini for providing photometry.
NB gratefully acknowledges funding from the Royal Society and European Research Council (grant  646928 ''Multi-Pop").
This research is also part of the project COSMIC-LAB (\url{http://www.cosmic-lab.eu}) funded by the European Research Council (under contract ERC-2010-AdG-267675).

\bibliographystyle{mn2e}
\bibliography{bibliography}

\begin{landscape}
\begin{tiny}
\begin{table}
 \centering
 \caption{Fe\,{\sc ii} atomic data and EWs presented by \citet{yong14} for the program stars.}
 \setlength{\tabcolsep}{0.1cm}
  \begin{tabular}{@{}lllrrrrrrrrrrrrrr@{}}
  \hline
\hline
\hline
Lambda &  E.P. &  $\log gf$ &  NR 37  &NR 38  &NR 47 &NR 58 & NR 60 &NR 76 & NR 77 & NR 81   &NR 99   &NR 124   &NR 132   &NR 207  &NR 254   &NR 378   \\
\AA	            &  eV   &                &    m\AA   &   m\AA         &m\AA   &m\AA   &m\AA   &m\AA   &m\AA   &m\AA   &m\AA   &m\AA    &m\AA    &m\AA    &m\AA     &m\AA     \\
\hline

4128.75 &  2.58 & $-$3.47 &   \ldots    & \ldots     & \ldots     &   58.6 (1) &  51.9  (1) &  \ldots     & \ldots     &  \ldots     &  \ldots     &  \ldots     &  \ldots       &  \ldots       &  \ldots       &  \ldots         \\
 4178.86 &  2.58 & $-$2.53 &   \ldots    & \ldots     & \ldots     &   80.1 (1) &  114.6 (-1) &  \ldots     & \ldots     &  \ldots     &  \ldots     &  \ldots     &	\ldots      &  \ldots       &   \ldots      &  \ldots         \\
 4416.82 &  2.78 & $-$2.43 &	 92.8 (1)& \ldots     & \ldots     &   95.9 (1) &  115.5 (-1) &    88.1 (-1) &  110.3 (1) &   105.4 (1) &  \ldots     &  \ldots     &	 116.7 (-1)  &  \ldots       &   \ldots      &  \ldots         \\
 4491.40 &  2.86 & $-$2.60 &	 75.9 (1)&   67.8 (1) & \ldots     &   71.6 (1) &  92.6  (1) &  \ldots     & \ldots     &    75.3 (1) &  \ldots     &  \ldots     &	 75.3  (-1)  &  \ldots       &   \ldots      &  \ldots         \\
 4508.29 &  2.85 & $-$2.31 &   \ldots    & \ldots     & \ldots     & \ldots     & \ldots     &    85.8 (-1) & \ldots     &    82.5 (1) &    91.4 (1) &   103.0 (1) &	  89.9 (-1)  &  \ldots       &     95.8 (-1)  &   108.2  (-1)    \\
 4508.30 &  2.86 & $-$2.28 &   \ldots    & \ldots     & \ldots     &   91.5 (1) & \ldots     &  \ldots     & \ldots     &  \ldots     &  \ldots     &  \ldots     &	\ldots      &  \ldots       &   \ldots      &  \ldots         \\
 4541.52 &  2.86 & $-$2.81 &	 62.7 (1)& \ldots     & \ldots     & \ldots     &  89.0  (1) &    62.4 (1) & \ldots     &  \ldots     &  \ldots     &  \ldots     &	 89.2  (-1)  &  \ldots       &   \ldots      &  \ldots         \\
 4555.89 &  2.83 & $-$2.18 &   \ldots    & \ldots     & \ldots     & \ldots     & \ldots     &    92.9 (-1) & \ldots     &  \ldots     &  \ldots     &  \ldots     &	\ldots      &  \ldots       &   \ldots      &  \ldots         \\
 4576.34 &  2.84 & $-$2.90 &	 68.1 (1)&   90.5 (1) & \ldots     &   72.7 (1) &  79.5  (1) &    62.5 (1) &   84.3 (1) &  \ldots     &  \ldots     &   75.0  (1) &	  90.2 (-1)  &  \ldots       &   \ldots      &    85.7  (1)   \\
 4582.84 &  2.84 & $-$3.09 &   \ldots    & \ldots     & \ldots     & \ldots     &  74.5  (1) &    57.0 (1) & \ldots     &    48.3 (1) &  \ldots     &  \ldots     &  \ldots       &  \ldots       &   \ldots      &  \ldots         \\
 4620.52 &  2.83 & $-$3.20 &	 47.6 (1)&   58.7 (1) &   44.4 (1) &   46.7 (1) &  59.6  (1) &    45.5 (1) &   42.8 (1) &    49.4 (1) &    48.6 (1) &   61.8  (1) &	  57.6 (1)  &  \ldots       &   \ldots      &    69.8  (1)   \\
 4833.19 &  2.66 & $-$4.62 &	 15.6 (1)& \ldots     & \ldots     & \ldots     & \ldots     &  \ldots     & \ldots     &    11.0 (-1) &  \ldots     &  \ldots     &	\ldots      &  \ldots       &   \ldots      &  \ldots         \\
 4840.00 &  2.68 & $-$4.74 &   \ldots    & \ldots     & \ldots     & \ldots     & \ldots     &  \ldots     &   22.2 (1) &  \ldots     &  \ldots     &  \ldots     &	\ldots      &  \ldots       &   \ldots      &  \ldots         \\
 4893.82 &  2.83 & $-$4.29 &   \ldots    & \ldots     & \ldots     &   20.1 (1) &  19.6  (1) &    16.2 (1) & \ldots     &  \ldots     &  \ldots     &  \ldots     &  \ldots       &  \ldots       &   \ldots      &  \ldots         \\
 4993.36 &  2.81 & $-$3.48 &   \ldots    & \ldots     &   40.4 (1) & \ldots     &  49.7  (1) &    32.4 (1) & \ldots     &    37.7 (1) &  \ldots     &   35.0  (1) &	  50.9 (1)  &    40.6  (1)  &    46.0  (1) &  \ldots         \\
 5100.66 &  2.81 & $-$4.16 &	 15.3 (-1)& \ldots     & \ldots     &   12.9 (-1) &  15.1  (-1) &    16.5 (1) & \ldots     &    17.7 (-1) &  \ldots     &  \ldots     &	 30.7  (1)  &  \ldots       &   \ldots     &  \ldots         \\
 5132.67 &  2.81 & $-$3.95 &	 18.1 (1)& \ldots     & \ldots     & \ldots     &  20.9  (1) &    14.8 (-1) & \ldots     &    14.9 (-1) &    16.4 (1) &  \ldots     &	 33.4  (1)  &  \ldots       &   \ldots     &  \ldots         \\
 5197.58 &  3.23 & $-$2.23 &	 78.6 (1)&   83.9 (1) & \ldots     &   79.9 (1) &  98.8  (1) &    77.7 (1) & \ldots     &    87.4 (1) &  \ldots     &  \ldots     &	 96.4  (-1)  &  \ldots       &   \ldots     &  \ldots         \\
 5234.63 &  3.22 & $-$2.22 &	 78.7 (1)&   78.0 (1) & \ldots     &   77.8 (1) &  93.3  (1) &    83.4 (-1) & \ldots     &    70.7 (1) &    86.8 (1) &  \ldots     &	 88.3  (-1)  &    81.0  (1)  &     87.8 (1) &   105.2   (-1)     \\
 5264.81 &  3.34 & $-$3.21 &	 40.2 (1)&   39.8 (1) &   33.1 (1) &   38.5 (1) &  43.6  (1) &    34.2 (1) & \ldots     &    30.2 (1) &  \ldots     &   45.6  (1) &	  42.9 (1)  &    47.2  (1)  &    51.9  ( 0) &  \ldots         \\
 5284.11 &  2.89 & $-$3.01 &   \ldots    & \ldots     & \ldots     & \ldots     & \ldots     &    51.2 (1) & \ldots     &    55.2 (1) &  \ldots     &    62.8 (1) &	  65.1 (1)  &  \ldots       &   \ldots     &  \ldots         \\
 5325.56 &  3.22 & $-$3.18 &	 28.7 (1)&   39.3 (1) & \ldots     &   40.4 (1) & \ldots     &    32.6 (1) & \ldots     &    37.4 (1) &  \ldots     &    43.0 (1) &	  41.1 (1)  &    43.8  (1)  &   \ldots     &  \ldots         \\
 5414.08 &  3.22 & $-$3.61 &   \ldots    & \ldots     & \ldots     & \ldots     &  18.4  (1) &  \ldots     & \ldots     &  \ldots     &  \ldots     &  \ldots     &  \ldots       &  \ldots       &   \ldots     &  \ldots         \\
 5425.26 &  3.20 & $-$3.27 &	 26.9 (1)&   29.7 (1) &   31.1 (1) &   28.7 (1) &  33.3  (1) &    32.0 (1) & \ldots     &    34.4 (1) &    31.7 (1) &   28.4  (1) &	  45.7 (1)  &    36.9  (1)  &    36.2  (1 ) &  \ldots         \\
 5525.13 &  3.27 & $-$4.00 &   \ldots    & \ldots     & \ldots     & \ldots     &  13.8  (-1) &  \ldots     & \ldots     &  \ldots     &  \ldots     &  \ldots     &  \ldots       &  \ldots       &   \ldots     &  \ldots         \\
 5534.85 &  3.25 & $-$2.75 &	 52.3 (1)&   56.0 (1) & \ldots     &   53.9 (1) &  60.5  (1) &    48.3 (1) & \ldots     &    50.8 (1) &  \ldots     &   57.5  (1) &	\ldots      &  \ldots       &   \ldots     &  \ldots        \\
 5991.38 &  3.15 & $-$3.56 &	 25.7 (1)&   21.9 (1) &   23.4 (1) &   19.0 (1) &  26.1  (1) &    27.6 (1) & \ldots     &    25.7 (1) &    22.8 (1) &   23.4  (1) &	  38.5 (1)  &    28.0  (1)  &    29.8  (1) &    39.9   (1)   \\
 6084.11 &  3.20 & $-$3.81 &	 13.7 (-1)&   15.0 (-1) & \ldots     &   13.8 (-1) & \ldots     &    13.0 (-1) & \ldots     &    15.9 (-1) &  \ldots     &  \ldots     &	  20.1 (1)  &    18.9  (-1)  &   \ldots     &    26.1   (1)   \\
 6113.33 &  3.22 & $-$4.13 &   \ldots    & \ldots     & \ldots     & \ldots     &  13.1  (-1) &  \ldots     &   14.6 (-1) &    15.4 (-1) &  \ldots     &  \ldots     &	 16.8  (-1)  &  \ldots       &   \ldots     &  \ldots         \\
 6149.25 &  3.89 & $-$2.73 &	 20.8 (1)& \ldots     & \ldots     &   18.7 (-1) & \ldots     &    24.3 (1) & \ldots     &    21.1 (-1) &    20.6 (1) &    22.9 (1) &	  28.0 (1)  &  \ldots       &   \ldots     &    39.9   (1)   \\
 6247.56 &  3.89 & $-$2.33 &	 34.5 (1)&   31.2 (1) &   28.2 (1) &   37.4 (1) &  42.0  (1) &    32.4 (1) &   40.0 (1) &    30.5 (1) &    40.0 (1) &   37.0  (1) &	  36.7 (1)  &  \ldots       &    44.5  (1) &    54.9   (1)   \\
 6369.46 &  2.89 & $-$4.21 &	 12.8 (-1)&   12.4 (-1) & \ldots     &   16.7 (-1) &  15.6  (-1) &    16.6 (-1) & \ldots     &    11.2 (-1) &  \ldots     &  \ldots     &	 20.0  (-1)  &    16.8  (-1)  &     23.2 (1) &  \ldots         \\
 6416.93 &  3.89 & $-$2.70 &   \ldots    & \ldots     & \ldots     &   21.3 (1) & \ldots     &    15.7 (-1) & \ldots     &  \ldots     &  \ldots     &  \ldots     &	\ldots      &  \ldots       &     32.1 (1) &  \ldots         \\
 6432.68 &  2.89 & $-$3.58 &	 31.0 (1)&   34.8 (1) &   29.3 (1) &   34.7 (1) &  38.3  (1) &    33.7 (1) &   31.0 (1) &    29.4 (1) &    34.7 (1) &   33.2  (1) &	  38.0 (1)  &    39.4  (1)  &    49.0  (1) &    53.6   (1)   \\
 6456.39 &  3.90 & $-$2.10 &	 46.1 (1)&   44.1 (1) &   35.0 (1) & \ldots     & \ldots     &  \ldots     &   48.5 (1) &    46.8 (1) &    45.8 (1) &  \ldots     &	  55.8 (1)  &    51.2  (1)  &     57.3 (1) &  \ldots         \\
 6516.08 &  2.89 & $-$3.38 &	 49.0 (1)&   46.9 (1) & \ldots     &   58.4 (1) &  56.5  (1) &    51.7 (1) & \ldots     &  \ldots     &    50.7 (1) &   49.1  (1) &	\ldots      &  \ldots       &    59.0  (1) &  \ldots         \\
 7711.72 &  3.90 & $-$2.54 &	 24.1 (-1)&   25.7 (1) & \ldots     &   21.9 (-1) &  28.3  (1) &  \ldots     &   35.0 (1) &  \ldots     &  \ldots     &  \ldots     &  \ldots       &  \ldots       &   \ldots      &  \ldots          \\
\hline
\end{tabular}
\flushleft{{\bf Notes:} the number in parenthesis is a flag indicating whether: the line is used in our analysis (1), the line is rejected as more than 3$\sigma$ away from the Fe\,{\sc ii} mean abundance (0), the line is rejected because outside the range of valid EWr values of Section~\ref{gspectro} (-1). Flags refer to the {\em hybrid} case of Section~\ref{gspectro}. \label{fe2_EW} }

\end{table}
\end{tiny}
\end{landscape}

\label{lastpage}
\end{document}